\documentstyle{article}
\def\beq{\begin{array}}
\def\eeq{\end{array}}
\begin{document}

\begin{center}
{\huge Strong interactions at all density}
\end{center}

\begin{center}

\subsection*{Vikram Soni}

National Physical Laboratory, K.S. Krishnan Marg,New Delhi 110012,India

 vsoni@del3.vsnl.net.in

\end{center}

\noindent \subsection*{Abstract}
 We show using Mean Field Theory that the phase diagram of QCD at finite 
density ($T=0$) is such that chiral symmetry remains spontaeously broken 
at all density!

\vfill
\begin{center}
Talk given at the Recontres Vietnam, Hanoi, August 2004
\end{center}
\vfill
\newpage

\section{Introduction}

We all know that nuclear matter and hadronic interactions at low energy
 are characterized by spontaneous chiral symmetry breaking ( SSB ), with the
 pion as the pseudo goldstone boson.   Further, the
 nature of the higher density quark matter state  is not clear - for example, is it in a
   spontaneous chiral symmetry broken state?

Can we find a theory that can describe both these domains? We present,
here, an Effective Chiral Intermediate Lagrangian, L,
that has quarks, gluons and a chiral multiplet of
 $[\vec\pi ,\sigma ]$ that flavor couples only to the quarks
 \cite{ref1,ref2,ref3,ref4,ref5,ref6}, which does so.

To begin with let us consider the two main features of the strong
interactions at low energy. These are i) that quarks  are
confined as hadrons and ii) chiral symmetry is spontaneously
broken ( SSB ) with the pion as an approximate Goldstone boson.
There is no specific reason that these two phenomena should
occur at an identical temperature scale , though QCD lattice
simulations show that for $ SU_2(L)\times SU_2(R)$ they are close.
The problem in giving an unequivocal answer to this question
is that we are yet to find a solution to the non perturbative aspects of
QCD.
Also, if the chiral symmetry restoration ( energy/temperature) scale was
lower
than the confinement scale we would expect hadrons to show parity
doubling below the confinement scale but above the chiral
SSB  scale .This is not seen in finite temperature lattice
simulations.

Actually , QCD can have multiple scales
\cite{ref5}.  Apart from, a confinement scale , a chiral symmetry restoration
scale we also have a 
compositeness scale for the pion. The above considerations suggest these scales
 are respectively in ascending order in energy  ( temperature ).

Let us consider the interacting fermi liquid of nucleons- nuclear matter. As
the baryon density is raised beyond overlap, we expect a transition to quark
 matter. An interesting question arises: Is the quark matter in a chiral
 SSB state with constituent quarks or is it , as usually assumed, in a
 chirally restored state with current quarks? As we will see the, L,
 given below,  answers to this question \cite{ref6,ref7}.

On the other hand there is some evidence for this intermediate
Chiral Lagrangian that has simultaneously no confinement but
chiral SSB. Such an effective Lagrangian has quarks, gluons and a
chiral multiplet of
 $[\vec\pi ,\sigma ]$ that flavor couples only to the quarks.

\begin{equation} 
L = - \frac{1}{4} G^a_{\mu v} G^a{\mu v}
 - \sum {\overline{\psi}} \left( D + g_y(\sigma +
i\gamma_5 \vec \tau \vec \pi)\right) \psi
- \frac{1}{2} (\partial
\mu \sigma)^2 - \frac{1}{2} (\partial \mu \vec \pi)^2
 - \frac{1}{2} \mu^2 (\sigma^2 + \vec \pi^2) - \frac{\lambda^2}{4}
(\sigma^2 + \vec \pi^2)^2 + \hbox{const}
\end{equation}

The masses of the scalar (PS) and fermions follow on the
minimization of the potentials above. This minimization yields

\begin{equation}
\qquad \mu^2 = - \lambda^2 <\sigma>^2
\end{equation}

It follows that

\begin{equation}
\qquad m_{\sigma}^2 = 2\lambda^2<\sigma>^2
\end{equation}

 This theory is an extension of QCD by additionally coupling the quarks
to a chiral multiplet , $( \vec\pi$ and $\sigma )$ \cite{ref1,
ref2,ref3,ref4}.

This Lagrangian has produced some interesting physics at the mean
field level \cite{ref4,ref8}

 (1) It provides a quark soliton model for the nucleon in which the nucleon
 is realized as a soliton with quarks being bound in a skyrmion
 configuration for the chiral field EVs
\cite{ ref1,ref4,ref8}.

 (i)Such a model gives a natural explanation for the 'Proton spin puzzle'.
 This is because the quarks in the background fields are in a spin, 'isospin'
 singlet state in which the quark spin operator averages to zero. On the
 collective quantization of this soliton to give states of good spin and
 isospin the quark spin operator acquires a small non zero contribution
 \cite{ref9}.

 (ii) Such a Lagrangian also seems to naturally produce the Gottfried sum
  rule \cite{ref10}.

 (iii) Such a nucleon can also yield from first principles (but with some
 drastic QCD evolution), structure functions for  the nucleon which is close
  to the experimental ones \cite{ref11}.

 (iv) In a finite temperature field theory such an effective Lagrangian also
  yields screening masses that match with those of a finite temperature QCD
  simulation with dynamical quarks \cite{ref12}.This work also does not show 
  any parity doubling for the hadronic states.

 (v) This Lagrangian also gives a consistent equation of state for strongly
 interacting matter at 'all' density  \cite {ref4,ref7,ref13}.

 This, L , has a single dimensional parameter, $f_\pi $ , that  is the pion decay
 constant, and three couplings, $ g_3 $ , the QCD coupling, $ g_y $, the yukawa coupling
 between quarks and mesons, that will be determined from the nucleon mass and
 the meson-meson coupling , $ \lambda $, which ,
  for this model, can be determined from meson meson scattering \cite{ref14}.
  No further
  phenomenological input will be used. As it stands, there is no confinement
  in this model, but may be dynamically generated as in QCD.

  Since this is posited as an effective L, we should have an appoximate idea of
  its range of validity . We find somewhat in analogy with the top quark (
  large yukawa coupling ) composite higgs picture, that we can get a compositeness
   scale for the scalars in this model by using RNG evolution. We find that the
   WF renormalisation for the scalars is inversely proportional to the running
   Yukawa coupling and thus naively vanishes when the yukawa coupling blows up.
   For our theory such a ballpark scale falls between 700 -800 Mev
   \cite{ref15}.

   An independent and quite general approach in setting a limit to the range
   of validity of non
   asmptotically free ( eg. yukawa theories ) is the vacuum instability
   to small length scale fluctuations ( or large momenta in quantum loop
   corrections ) that plagues
   these theories, discovered and analytically proved by one of us \cite{ref16}.
   The scale at which this occurs is of the same order as above. This is not
   very surprising since it is connected to non AF character of the yukawa
   coupling \cite{ref16,ref17}. This underscores the impossibility of doing loop
   calculations, unless
   we introduce a cut off, even though our ,L , is renormalizable!

 Given these facts we shall use this theory  at the Mean Field level to look
 at different phases of this field theory.

    For this ,L, the nucleon is
    a colour singlet bound state of three valence quarks in a skyrme background,
     in a linear sigma model. Fixing the nucleon mass to be the observed one
     fixes the yukawa coupling. A generally accepted value for , $ g_y =5.4 $
     \cite {ref1}. This is
     when the $ \pi$  and $ \sigma $ fields are varied independently, without
     making any ansatz,
     and the quark soliton so obtained is projected to give a nucleon with good spin
     and isospin.

     We now have a determination for all parameters of our , L:
     $ f_\pi, g_3 $ and $ g_y $ , leaving only one parameter, $ \lambda $ to be set.
     As we shall show  this can be determined from low energy meson -meson
     scattering data.

     Recently, Schechter et al \cite{ref14} made a fit to scalar scattering data
     to see how it may be fitted with increasing  $\sqrt{s} $ ,
     using chiral perturbation theory and sevral resonances. They further looked at this
     channel using just a linear sigma model. Their results indicate that for
      $\sqrt{s} < 800 Mev $, a reasonable fit to the data can be made using the
      linear sigma model with a sigma mass above  but  close to 800 Mev. 
      This sets the value of $ \lambda $ .

     This completes the  determination of all the parameters of our, L,
     which is able to describe both nucleon and quark phases of dense matter.

   \section{Phases at finite density}

   We now move to the main theme of this work , which is to look at ground state
   of strongly interacting matter at finite density. In our model all the phases
   considered are characterised simply by different patterns of chiral spontaneous
    symmetry breaking ( chiral SSB ). As we shall see  i) the nucleonic low
    density phase is characterised
     by the pattern of SSB in skyrmions \cite{ref13} ii) the quark matter
      Lee Wick phase has space uniform
     SSB, with only the sigma field expectation value. As density goes up
     the chiral symmetry restoration transition occurs and we get chirally
     restored quark matter ( CRQM ) \cite{ref7,ref13}. and iii) the pion
     condensed phase ( PC ) is characterised by a stationary wave, with
     a space dependent periodic variation in
     the sigma and neutral pion field expectation values. In contrast to case
     ii), in this case chiral symmetry restoration does not occur. This state is
     always lower in energy than the CRQM above. For this we refer the
     reader to \cite{ref6,ref7}
     
   \section {The phase diagram and the equation of state}

    The starting point for the phase diagram of QCD at finite density is as such:
   At vey low density we know that there is  chiral SSB, with the pion as
 the Goldstone boson - this breaks chiral symmetry spontaneously, leaving colour
symmetry unbroken. At very high density we have a colour SC pairing instability
for the quarks and of the many pairings invesigated the CFL ( colour flavour
locked ) pairing is favoured - this
spontaneously breaks colour and chiral symmetry. In between these limits the issue
of the ground state is open.

We approach this problem, from the low density end, by the effective lagrangian, L,that can access
 both , nuclear and quark matter states, with or without chiral SSB. We have
 argued that this, L, has validity upto energy scales of about 800 Mev.
 The advantage of this , L , is that it has
only 3 couplings and no other free parameters. These can all be determined from 
experiment. However, this, L, has only colour singlet scalar/psedoscalar meson
fields, but no colour scalar fields. It is thus good for investigating chiral
condensates but not colour condensates.

If we start at the high density end, as done by the several authors who
practice diquark
colour superconductivity, one knows the ground state at very high density,
when asymptotically free QCD makes calculation reliable.
However, for intermediate density  phenomenological four fermion interactions have to 
be introduced. This is equivalent to introducing a given value for the gap
parameter, $ \Delta $ \cite{ref18}. Also, another phenomenological input is
 the bag parameter , B. Uncertainty attends these parameters.
It is not clear how this ground state transits into the low density chiral
SSB state.

    We start with saturation nuclear matter at nuclear density and this
   persists till the nuclear matter gets squeezed into quark matter at
   moderately higher density, when a pion condensed state takes over \cite{ref3}.
  In all these phases ,as we have found, chiral symmetry is spontaneously broken -
  though, the patterns of symmetry breaking keep changing with baryon density.
 Since we have not investigated all other possible condensates, 
   we cannot vouch for our neutral pion condensate being the best ground state.

   It is, however, to be noted from the lecture notes of Baym \cite{ref19}
the neutral pion condensate is preferred over the charged pion condensate for
 charge neutral nuclear matter, in the non relativistic limit ( particularly
 if we put , $ g_A = 1 $,). We find that this is also the case for charge
 neutral quark matter as well.
   It is worth pointing out that all these states , that have lower
   energy than the chirally restored CRQM state, are chiral symmetry broken
   states.

   At even higher density the most likely state is a diquark condensate - a
   colour flavour locked (  CFL ) \cite{ref18} state - which can persist till
    arbitrarily
   high density. Such a ground state spontaneously breaks both chiral and
   colour symmetry.
   Diquark condensates are unlikely at moderately high density as they depend
   on the quark density of states and so the pion condensate is most likely
   at such densities.

   There is an important issue that arises here and that is the comparision
    between the diquark condensate state and the pion condensed state.
    In this case the starting point is a chiral symmetric four fermion
    interaction which can accommodate both chiral ( quark-antiquark colour
     singlet ) condensates and diquark condensates. Such a
    comparision has already been done by Sadzikowski\cite{ref20} in the
    context of a NJL chiral symmetric model, for the case of 2 flavours - $ SU(2)_L \times
    SU(2)_R $. What is done is at the level of mean field
    theory. The NJL model has four fermion interactions in terms of the
    quark bilinears corresponding to the $ \sigma$ and $\pi $  field quantum
     numbers, with a common dimensional coupling , G. If we are interested in
     a ground state carrying sigma and/or pion condensates we can replace these
     quark bilinears by the  corresponding  $ \sigma$ and $ \pi $ EV's in
     the MFT. This yields the ground state energy of the space uniform SSB
     and the PC states.
     Alternatively, this NJL can be mapped to our linear sigma model and
     the ground state thereof, which has been considered in \cite{ref6}.

     To get to the diquark condensate state  we have to Fierz
     transform this NJL chiral , L, and look for its projection into the diquark
     condensate channel and follow the same procedure of MFT. This projection
     gives the diquark condensate, L, used by \cite{ref18,ref20}, with a four 
      fermion coupling,  $ G^{'}= G/4 $\cite{ref20}.

     i) Working with simultaneous MF condensates, corresponding to  space uniform 
    chiral SSB ( which generates  a spontaneous mass for the quarks (no PC) ) and 
    diquark SC ( which gives rise to colour SC ), they \cite{ref20} find that for
     the
     above value of the two
     G's, that follow from the Fierz projection, for the two condensates, the
    chiral condensate is always the preferred state up till the limit of
    validity of this model.

    However, they find that with arbitrary values of G's for the two
    condensates, it is possible to have a phase transition from
    a space uniform chiral SSB  to a diquark condensate at some large  density.
     In paricular for , $ G^{'}= G $,  they find that that from a pure chiral
     SSB state there is a continuous phase transition to a mixed chiral SSB
     and diquark condensate state at a quark chemical potential of about
     0.33 Gev.This is followed by a first order phase transition at slightly
     higher $ \mu $, to also a dominantly diquark condensate mixed state.

   This state then evolves to an almost entirely diquark condensate.

ii) Till now we have not cosidered the PC state.The PC state always has lower
energy 
( free energy ) than the uniform chiral SSB state. It would then be reasonable to 
expect that the PC state and not the uniform state would be the preferred state of 
chiral SSB. Since chiral SSB persists till much higher density ( $ \mu $ ) in this 
state we expect that with the inclusion of this state the transition to the colour SC 
state will be pushed to higher density ( $ \mu $ ).

A following work by the same author addresses this matter by considering simultaneous 
MF condensates of the space uniform chiral SSB , PC, and the the diquark condensate 
\cite{ref21}.
It is found that the first order transition does shift to higher desity ($ \mu $ ).
This is for the case  $ G^{'} = G/2 $. Furthermore , in this case it seems after the 
first order transition occurs, the chiral SSB order parameter , M, goes to zero, 
indicating that this state has no chiral SSB but only colour SSB.
Of course , this is so as these works deal with the 2 flavour case - where the clour 
diquark condensate is a chiral singlet.

Realistically, we must consider 3 flavours, since the quark chemical potential is
much greater than the strange quark mass. In this case we are very likely to have the 
CFL state as the lowest energy state. The criterion for this is given in 
\cite{ref22} and is , $ \Delta > \frac{m_s^2}{4\mu} $, which is easily satisfied.
Furthermore,in this case the diquark condensate is a colour, flavour condensate which 
has both chiral SSB and colour SSB, albeit in a manner different to the the PC.

The deciding question is then what effective , L, is to be used. There is no
derivation of the exact effective, L, from QCD at arbitrary scales.There are
many four fermion type ( or higher order ) interactions derived or rather
motivated from various points of view, eg instantons, gluon exchange etc.
However, there is no unique, L, that is derived as such. It is therefore
necessary to have an argument to decide this issue.

 Since the NJL chiral , L,  identifies  with our linear sigma model, and this
 latter model is what we have used as a valid model till centre of mass
 energies/scales of less than 800 Mev, the right procedure is the take this
 model to describe physics upto this scale. In this case, as we have argued,
 the PC is the preferred ground state, till the scale of validity of our
 effective L  ( this is for  $ G^{'} =  G/4 $ or even if we relax this to
 larger $ G^{'}$ ) or we have a 3 flavour PC followed by a CFL state , with
 increasing density.

     With increasing baryon density we then find the following heirarchy. At
     nuclear density and above we have nuclear matter with chiral SSB, followed
     by the pion condensed quark matter, again with chiral SSB, albeit with a
     different realization and finally a transition to the diquark CFL state
     which also has chiral SSB  ( and colour SSB ), with yet another
     realization. A point to note is that CRQM or free fermi seas are
     unstable. To one's surprise at zero temperature, at any finite density
     chiral symmetry is never restored!


\newpage

\noindent \subsection*{References}



\begin{enumerate}

\bibitem {ref1} (a) S. Kahana, G. Ripka and V. Soni , Nucl. Phys. A415,
351 (1984),

(b) M. C. Birse and M.K. Banerjee, Phys. lett. 134B, 284, (1984).

  

\bibitem {ref2} A. Manohar and H. Georgi, Nucl. Phys.B234,  203 (1984)  .

\bibitem {ref3} V. Soni, Mod. Phys. Lett. A, Vol.11, 331 (1996),

\bibitem {ref4} V. Soni, The nucleon and strongly interacting matter',Invited
 talk at DAE Symposium in Nuclear Physics Bombay , Dec  1992 and references
 therein.

\bibitem {ref5}  V Soni , M.S. Modgil and D. Sahdev, Phys. Lett. B 552, 165
 ( 2003 ).

\bibitem {ref6}  V Soni and D. Bhattacharya, Phys. Rev. D 69, 074001
 ( 2004 ).



\bibitem {ref7} M. Kutschera, W. Broniowski and A. Kotlorz,Nucl. Phys. A516
  566 (1990).

\bibitem {ref8} M. C. Birse, Soliton Models in Nuclear Physics, Progress  in
 Particle and Nuclear Physics, Vol. 25 , (1991) 1 ,and references therein.


\bibitem {ref9}  see for example, R.Johnson, N. W. Park, J. Schechter and 
V. Soni and H. Weigel, Phys. Rev. D42 , 2998 (1990), J. Stern and G. Clement,
 Mod Phys. Lett, A3, 1657 (1988) .



\bibitem {ref10} see for example, J. Stern and G. Clement , Phys. Lett. B 264,
 426  (1991),
E.J. Eichten,I Hinchcliffe and C.Quigg ,Fermilab-Pub 91/272 - T.



\bibitem {ref11} D. Diakonov, V. Petrov, P. Pobylitsa, M Polyakov and C. Weiss
, Nucl. Phys. B480, 341 (1996), Phys. Rev. D56, 4069 (1997).



\bibitem {ref12} A. Gocksch , Phys. Rev. Lett. 67 , 1701 (1991).



\bibitem {ref13} V. Soni, Phys. Lett. 152B ,231 (1985).



\bibitem {ref14} A. Abdel-Rehim, D. Black, A.H. Fairbroz, S. Nazri and
 J. Schechter, Phys. Rev D68, 013008 ( 2003 ).



\bibitem {ref15} V. Soni ( unpublished )



\bibitem {ref16} V. Soni, Phys. Lett. B183,91 ( 1987 ) , Phys. Lett. B195, 569
( 1987 ). 

\bibitem {ref17} see also  related work, R. Perry, Phys. Lett B199, 489
( 1987), 
G. Ripka and S. Kahana, Phys. Rev. D 36 1233 ( 1987 ), W. Boniowski and M. 
Kutschera , Phys.lett. B 234 449 (1990 ).

 

\bibitem {ref18}  M. Alford , K. Rajgopal and F. Wilczek, Phys. Lett B
 422, 247 (1998) ,Nucl. Phys. B537, 443 (1999). 

\bibitem {ref19}
 G. Baym in NORDITA lectures, 'Neutron Stars and the Properties of Matter
 at high density', (1977).
 see also ,G.Baym and S. Chin, Phys. Lett. 62B, 241 (1976)

\bibitem {ref20} M. Sadzikowski, Mod. Phys. Lett. A Vol.16, No. 17 1129 (2001)
\bibitem {ref21} M. Sadzikowski, hep-ph/ 0210065v2 (2003)


\bibitem {ref22}  M. Alford , K. Rajgopal,S. Reddy and F. Wilczek, Phys. Rev
D 64, 074017 ( 2001 )



\end{enumerate}
\end{document}